\documentstyle[12pt]{article} 
\normalbaselines
\begin{document}
\bibliographystyle{unsrt} 

\vbox {\vspace{6mm}}

\begin{center}
{\large \bf MAXIMUM PREDICTIVE POWER AND THE SUPERPOSITION
PRINCIPLE} \footnote{This is a slightly revised version of a paper which 
appeared in {\it Int. J. Theor. Physics {\bf 33}, 171-178 (1994).}}
\\[7mm]
Johann Summhammer\\ 
{\it Atominstitut der \"Osterreichischen Universit\"aten\\ Stadionallee 2,
A-1020 Vienna, AUSTRIA}\\[5mm]
\end{center}

\vspace{2mm}

\begin{abstract}
In quantum physics the direct observables are probabilities of events. We ask, how observed
probabilities must be combined to achieve what we call maximum predictive power. 
According
to this concept the accuracy of a prediction must only depend on the number of runs whose 
data
serve as input for the prediction. We transform each probability to an associated 
variable whose uncertainty interval depends only on the amount of data and strictly decreases 
with it. We find that for a probability which is a function of two other probabilities maximum
predictive power is achieved when linearly summing their associated variables and 
transforming back to a probability. This recovers the quantum mechanical superposition
 principle.
\end{abstract}

\section{Introduction} 
Quantum theory is not yet understood as well as e.g. 
classical mechanics or special relativity. Classical mechanics coincides well 
with our intuition and so is rarely questioned. Special relativity runs 
counter to our immediate insight, but can easily be derived by assuming 
constancy of the speed of light for every observer. And that assumption may be made 
plausible
by epistemological 
arguments \cite{1}. Quantum theory on the other hand demands two 
premises. First, it wants us to give up determinism for the sake of a 
probabilistic view. In fact, this seems unavoidable in a fundamental theory 
of prediction, because any communicable observation can be decomposed into a 
finite number of bits. So predictions therefrom always have limited accuracy, and probability
enters naturally. 
More disturbing is the second premise: Quantum theory wants us to give up the 
sum rule of probabilities by requiring interference instead. However, the sum 
rule is deeply ingrained in our thought, because of its roots in counting and the definition of 
sets:
Define sets with no common elements, then define the set which joins them all. 
The number of elements in this latter set is just the sum of the elements of 
the individual sets. When deriving the notion of probability from the relative 
frequency of events we are thus immediately led to the sum rule, such that any 
other rule appears inconceivable. And this may be the reason why we have 
difficulties accepting the quantum theoretical rule, where probabilities are 
summed by calculating the square of the sum of the complex square roots of the 
probabilities. In this situation two views are possible. We may either consider 
the quantum theoretical rule as a peculiarity of nature. Or, we may conjecture 
that the quantum theoretical rule has something to do with how we organize 
data from observations into quantities that are physically meaningful to us. 
We want to adopt the latter position. Therefore we seek to establish a grasp of 
the quantum theoretical rule with the general idea in mind that, given the 
probabilistic paradigm, there may exist an optimal strategy of prediction, 
quite independent of traditional physical concepts, but resting on what one 
can deduce from a given amount of information. We will formulate elements of 
such a strategy with the aim of achieving maximum predictive power.

\section{Representing Knowledge from Probabilistic Data}
Any investigative endeavour rests upon one natural assumption: More data from 
observations will lead to better knowledge of the situation at hand. Let us see 
whether this holds in quantum experiments. The data are relative frequencies of 
events. From these we deduce probabilities from which in turn we derive the 
magnitudes of physical quantities. As an example take an experiment with two 
detectors, where a click is registered in either the one or the other. 
(We exclude simultaneous clicks for the moment.) Here, only one probability is 
measurable, e.g. the probablity $p_1$ of a click in detector 1. After N runs we 
have $n_1$ counts in detector 1 and $n_2$ counts in detector 2, 
with $n_1+n_2=N$. The probability $p_1$ can thus be estimated as
\begin{equation} p_1 = \frac{n_1}{N} \end{equation}
with the uncertainty interval \cite{2}
\begin{equation} \Delta p_1 = \sqrt{\frac{p_1(1-p_1)}{N}} . 
\end{equation}
From $p_1$ the physical quantity $\chi(p_1)$ is derived. Its uncertainty interval is
\begin{equation}
 \Delta\chi = \left| \frac{\partial\chi}{\partial p_1} 
\right|\Delta p_1 = 
 \left| \frac{ \partial \chi}{\partial p_1} \right| 
 \sqrt{\frac {p_1(1-p_1)}{N}} . 
\end{equation}
The accuracy of $\chi$ is given by the inverse of $\Delta \chi$. 
With the above
assumption we expect it to increase with each additional run, because we get 
additional data.
Therefore, for any $N$, we expect
\begin{equation} \Delta \chi (N+1) < \Delta \chi (N) . \end{equation}
However, this inequality cannot be true for an arbitrary function $\chi(p_1)$. 
In general
$\Delta\chi$ will fluctuate and only decrease on the average with increasing 
$N$. To see this take a
theory A which relates physical quantity and probability by $\chi_A=p_1$. In 
an experiment of
$N=100$ runs and $n_1=90$ we get: $\Delta\chi_A(100)=.030$. By taking into 
account the data from one
additional run, where detector 2 happened to click, we have $\Delta\chi_A(101)
=.031$. The differences
may appear marginal, but nevertheless the accuracy of our estimate for $\chi_A$
 has decreased
although we incorporated additional data. So our original assumption does not 
hold. This is
worrisome as it implies that a prediction based on a measurement of $\chi_A$ 
may be more accurate
if the data of the last run are not included. Let us contrast this to theory B, 
which connects
physical quantity and probability by $\chi_B={p_1}^6$. With $N$ and $n_1$ as 
before we have $\Delta\chi_B(100)=.106$.
Incorporation of the data from the additional run leads to $\Delta\chi_B(101)=.104$. 
Now we obviously
don't question the value of the last run, as the accuracy of our estimate has 
increased.

The lesson to be learnt from the two examples is that the specific functional
dependence of a physical quantity on the probability (or several probabilities 
if it is derived
from a variety of experiments) determines whether our knowledge about the 
physical quantity
will increase with additional experimental data, and that this also applies to 
the accuracy of
our predictions. This raises the question what quantities we should be 
interested in to make
sure that we get to know them more accurately by doing more experiments. 
From a statistical
point of view the answer is straightforward: choose variables whose uncertainty 
interval
strictly decreases, and simply {\it define} them as physical. And from a 
physical 
point of view?
Coming from classical physics we may have a problem, as concepts like mass, 
distance,
angular momentum, energy, etc. are suggested as candidates for physical 
quantities. But when
coming from the phenomenology of quantum physics, where all we ever get from 
nature is
random clicks and count rates, a definition of physical quantities according 
to statistical
criteria may seem more reasonable, {\it simply because there is no other guideline 
as to which
random variables should be considered physical.}

Pursuing this line of thought we want to express experimental results by 
random
variables whose uncertainty interval strictly decreases with more data. 
When using them in
predictions, which are also expressed by variables with this property, 
predictions should
automatically become more accurate with more data input. Now a few trials will 
show that
there are many functions $\chi(p_1)$ whose uncertainty interval decreases with 
increasing $N$ (eq.(3)).
We want to choose the one with maximum predictive power. The meaning of this 
term
becomes clear when realizing that in general $\Delta\chi$ depends on $N$ and 
on $n_1$ (via $p_1$). These two
numbers have a very different status. The number of runs, $N$, is controlled 
by the
experimenter, while the number of clicks, $n_1$, is solely due to nature. 
Maximum predictive
power then means to eliminate nature's influence on $\Delta\chi$. For then we 
can know $\Delta\chi$ {\it even
before} having done any experimental runs, simply upon deciding how many we 
will do. From
eq.(3) we thus get
\begin{equation} \sqrt{N} \Delta\chi = 
\left|\frac{\partial\chi}{\partial p_1}
\right|\sqrt{p_1(1-p_1)} = constant, \end{equation}
which results in 
\begin{equation} \chi = C\arcsin( 2p_1 - 1 ) + D \end{equation}
where C and D are real constants. The inverse is
\begin{equation} p_1 = \frac{1 + \sin( \frac{\chi - D}{C} )}{2} , 
\end{equation} 
showing that the probability is periodic in $\chi$. Aside from the linear 
transformations provided
by $C$ and $D$ any other smooth function $\alpha(\chi)$ in real or complex 
spaces will also fulfill
requirement (5) when equally sized intervals in $\chi$ correspond to equal 
line lengths along the
curve $\alpha(\chi)$. One particular curve is
\begin{equation} \alpha(\chi) = \sin(\frac{\chi}{2}) 
e^{i\frac{\chi}{2}} ,
\end{equation}
which is a circle in the complex plane with center at $i/2$. It exhibits the 
property $\left|\alpha\right|^2=p_1$ known from quantum theory. But note, that 
for instance the
function $\beta = 
\sin(\chi/2)$ does not
fulfill the requirement that the accuracy only depend on $N$. Therefore the 
complex phase
factor in eq.(8) is necessary \cite{3}\cite{4}.

\section{Distinguishability}
We have now found a unique transformation from a probability to another class of variables
exemplified by $\chi$ in eq.(6). These unique variables always become better known with
additional data. But can they be considered physical? We should first clarify what a physical
variable is. A physical variable can assume different numerical values, where each value
should not only imply a different physical situation, but should most of all lead to a different
measurement result in a properly designed experiment. Within the probabilistic paradigm two
measurement results are different when their uncertainty intervals don't overlap. This can be
used to define a variable which counts the principally distinguishable results of the
measurement of a probability. Comparison of that variable to our quantity $\chi$ should tell 
us
how much $\chi$ must change from a given value before this can be noticed in an 
experiment.
Following Wootters and Wheeler \cite{5}\cite{6} the variable $\theta$ counting 
the statistically
distinguishable
results at detector 1 in $N$ runs of our above example is given by
\begin{equation}
 \theta(n_1) = \int_0^{p_1(n_1)} \frac{dp}{\Delta p(p)} =
 \sqrt{N} \left[ \arcsin ( 2p_1 - 1 ) + \frac{\pi}{2} \right]_{ {p_1} =
 \frac{n_1}{N} }
\end{equation}
where $\Delta p$ is defined as in eq.(2). When dividing $\theta$ by $\sqrt{N}$ 
it becomes identical to $\chi$ when in
eq.(6) we set $C=1$ and $D=\frac{\pi}{2}$. This illuminates the meaning of 
$\chi$: It is a continuous variable
associated with a probability, with the particular property that anywhere in its domain an
interval of fixed width corresponds to an equal number of measurement results 
distinguishable
in a given number of runs. With Occam's dictum of not introducing more entities than are
necessary for the description of the subject matter under investigation, $\chi$ 
would be {\it the} choice
for representing physical situations and can rightly be called physical.

\section{A Simple Prediction: The Superposition Principle}
Now we return to our aim of finding a strategy for maximum predictive power.
We want to
see whether the unique class of variables represented by $\chi$ indicates a way
beyond representing data and perhaps affords special predictions. For the sake of
concreteness we
think of the double slit experiment. A particle can reach the detector by two
different routes.
We measure the probabilty that it hits the detector via the left route, $p_L$, by
blocking the right
slit. In $L$ runs we get $n_L$ counts. In the measurement of the probability with
only the right
path available, $p_R$, we get $n_R$ counts in $R$ runs. From these data we want
to make a prediction
about the probability $p_{tot}$, when both paths are open. Therefore we make the
hypotheses that
$p_{tot}$ is a function of $p_R$ and $p_L$. What can we say about the function
$p_{tot}(p_L,p_R)$ when we
demand maximum predictive power from it? This question is answered by
reformulating the
problem in terms of the associated variables $\chi_L$, $\chi_R$ and $\chi_{tot}$,
which we derive according to
eq.(6) by setting $C=1$ and $D=\frac{\pi}{2}$. The function
$\chi_{tot}(\chi_L,\chi_R)$ must be such that a prediction for
$\chi_{tot}$ has an uncertainty interval $\delta \chi_{tot}$, which only depends
on the number of runs, $L$ and $R$, and
decreases with both of them. (We use the symbol $\delta \chi_{tot}$ to indicate
that it is not derived from a
measurement of $p_{tot}$, but from other measurements from which we want to
predict $p_{tot}$.) In this
way we can predict the accuracy of $\chi_{tot}$ by only deciding the number of
runs, $L$ and $R$. No
actual measurements need to have been done. Because of
\begin{equation}
 \delta \chi_{tot} = \sqrt{ \left|\frac{\partial \chi_{tot}}
 {\partial \chi_L} \right|^2 \frac{1}{L} + \left|\frac{\partial 
\chi_{tot}}
 {\partial \chi_R} \right|^2 \frac{1}{R} }
\end{equation}
maximum predictive power is achieved when
\begin{equation}
 \left|\frac{\partial \chi_{tot}} {\partial \chi_j} \right| =
 constant,\mbox{\hspace{0.5cm}}    j=L,R.
\end{equation}
We want to have a real function $\chi_{tot}(\chi_L,\chi_R)$, and therefore we 
get
\begin{equation} \chi_{tot} = a \chi_L + b \chi_R + c, \end{equation}
where $a$, $b$ and $c$ are real constants. Furthermore we must have $c=0$ and
the magnitude of both
$a$ and $b$ equal to $1$, when we wish to have $\chi_{tot}$ equivalent to
$\chi_R$ or to $\chi_L$ when either the one
or
the other path is blocked. So there is an ambiguity of sign with $a$ and $b$.
When rewriting this
in terms of the probability we get
\begin{equation} p_{tot} = \sin^2( \frac{\chi_L \pm \chi_R}{2} ). 
\end{equation}
This does not look like the sum rule of probability theory. Only for $p_L+p_R=1$
does it coincide
with it. We may therefore conclude that the sum rule of probability theory does
not afford
maximum predictive power. But neither does eq.(13) look like the quantum
mechanical
superposition principle. However, this should not be surprising because our input
were just
two real valued numbers, $\chi_L$ and $\chi_R$, from which we demanded to
derive another real valued
number. A general phase as is provided in quantum theory could thus not be
incorporated.
But let us see what we get with complex representatives of the associated variables
of
probabilities. We take $\alpha(\chi)$ from eq.(8). Again we define in an equivalent
manner $\alpha_L$, $\alpha_R$ and
$\alpha_{tot}$. From $p_L$ we have for instance (from (8) and (7) with $C=1$
and $D=\frac{\pi}{2}$)
\begin{equation} \alpha_L = \sqrt{p_L} \left( \sqrt{p_L} + i\sqrt{1-p_L} 
\right)
\end{equation}
and
\begin{equation} \Delta \alpha_L = \left| \frac {\partial \alpha_L} 
{\partial p_L}
\right| \Delta p_L = \frac {1} {2 \sqrt{L} }.  \end{equation}
If we postulate a relationship $\alpha_{tot}(\alpha_R,\alpha_L)$ according to
maximum predictive power we expect
the predicted uncertainty interval $\delta\alpha_{tot}$ to be independent of
$\alpha_L$ and $\alpha_R$ and to decrease with
increasing number of runs, $L$ and $R$. Analogous to (11) we must
have
\begin{equation}
 \left| \frac {\partial \alpha_{tot}} {\partial \alpha_j} \right| =
 constant, \mbox{\hspace{0.5cm}}    j=L,R,
\end{equation}
yielding
\begin{equation} \alpha_{tot} = s\alpha_L + t\alpha_R + u, \end{equation}
where $s$, $t$, and $u$ are complex constants. Now $u$ must vanish and $s$ and
$t$ must both be
unimodular when $p_{tot}$ is to be equivalent to either $p_L$ or $p_R$ when the
one or the other route is
blocked. We then obtain
\begin{equation} p_{tot} = \left| \alpha_{tot} \right|^2 = \left| 
s\alpha_L +
t\alpha_R \right|^2 = p_L + p_R + 2\sqrt{p_L p_R} \cos(\phi), 
\end{equation}
where $\phi$ is an arbitrary phase factor containing the phases of $s$ and $t$.
This is exactly the
quantum mechanical superposition principle. What is striking is that with a theory
of
maximum predictive power we can obtain the general form of this principle, but
cannot at all
predict $p_{tot}$ even when we have measured $p_L$ and $p_R$, because of the
unknown phase $\phi$. So we are lead to postulate $\phi$ as a new measurable 
quantity in this experiment.

\section{Conclusion}
We have tried to obtain insight into the quantum mechanical superposition
principle and set
out with the idea that it might follow from a most natural assumption of
experimental science:
more data should provide a more accurate representation of
the matter
under investigation and afford more accurate predictions. From
this we
defined the concept of maximum predictive power which demands laws to
be such that the uncertainty of a prediction is solely dependent on the number of
experiments
on which the prediction is based, and not on the specific outcomes of these
experiments.
Applying this to the observation of two probabilities and to possible predictions
about a third
probability therefrom, we arrived at the quantum mechanical superposition principle. Our 
result
suggests nature's law to be such that from more observations more accurate predictions
must be derivable.

\section{Acknowledgments}
I thank the Austrian Science Foundation (FFW) for financial support of ion double slit
experiments (Project P8781-PHY) whose analysis led to this paper.


\begin{thebibliography}{99}

\bibitem{1} G. J. Whitrow, {\it The Natural Philosophy of Time, 2nd Ed.}, 
(Clarendon Press, Oxford, 1984).
\bibitem{2} The uncertainty interval is derived from Chebyshev's inequality, 
e.g. William Feller, {\it An Introduction to Probability Theory and its 
Applications, 3rd Ed.}, (Wiley and Sons, New York, 1968), p.233. For reasons of 
simplicity we are using only the approximate form valid for large $N$.
\bibitem{3} More details and variables that can be used to represent several
probabilities can be found in: J. Summhammer, Found. Phys. Lett. {\bf 1}, 113 
(1988).
\bibitem{4} The statistical properties of $\chi$ are analyzed in: J. Summhammer, 
Phys. Lett. {\bf A136}, 183 (1989).
\bibitem{5} W. K. Wootters, Phys. Rev. {\bf D23}, 357 (1981).
\bibitem{6} J. A. Wheeler, Int. J. Theor. Phys. {\bf 21}, 557 (1982).


\end {thebibliography}
\end{document}